# A National Discovery Cloud: Preparing the US for Global Competitiveness in the New Era of 21st Century Digital Transformation

*Ian Foster (Argonne National Laboratory and University of Chicago), Daniel Lopresti (Lehigh University), Bill Gropp (University of Illinois Urbana-Champaign), Mark D. Hill (Microsoft and University of Wisconsin), Katie Schuman (Oak Ridge National Laboratory)*

The nature of computation and its role in our lives have been transformed in the past two decades by three remarkable developments: the emergence of **public cloud utilities** as a new computing platform; the ability to extract information from **enormous quantities of data** via machine learning; and the emergence of **computational simulation** as a research method on par with experimental science. Each development has major implications for how societies function and compete; together, they represent a change in technological foundations of society as profound as the telegraph or electrification.

Societies that embrace these changes will lead in the 21st Century; those that do not, will decline in prosperity and influence. Nowhere is this stark choice more evident than in research and education, the two sectors that produce the innovations that power the future and prepare a workforce able to exploit those innovations, respectively.

In this article, we introduce these developments and suggest steps that the US government might take to prepare the research and education system for its implications.

**The three pillars of the new computational fabric**

Cloud platform: In the 1960s, computing was limited to the few rich corporations and government laboratories who could afford a multimillion-dollar mainframe. From the 1980s onward, the PC provided a simple, low-cost platform that allowed many more people to write programs, spurring an innovation explosion that changed everything, as exemplified by the World Wide Web. Today, those with access to cloud platforms via an Internet connection can author persistent, reliable, secure services that scale to reach thousands (or, if needed, millions) of users. Want to allow anyone to run a Jupyter notebook[1]? Operate a research data management service for thousands of research institutions[2]? 20 years ago, such tasks would have required a tremendous amount of engineering. Today, cloud capabilities mean that you can create them in weeks and scale them in an afternoon. Groups who have access to cloud platforms have a large and growing competitive advantage over those who do not.

---

[1] https://mybinder.org
[2] https://globus.org

Data-driven discovery: Advances in sensors and storage mean that vast fields of human endeavor, from transportation to environmental monitoring, that were previously data starved, are now data rich. Meanwhile new computer architectures and machine learning methods enable the extraction of patterns and rules from vast data collections—the sometimes over-hyped but highly consequential AI revolution[3]. These developments are driving a great increase in both the quantity and value of data (as well as major new privacy and security concerns: see below). Groups with access to such datasets have a large and growing competitive advantage over those who do not.

Simulation: Also of vital importance is the emergence of computational modeling and simulation methods that, for example, allow aircraft to be designed, constructed, and flown without extensive flight tests, and new materials and new drugs to be designed in months rather years. Enabled by immensely powerful supercomputers, these methods allow simulation to proceed as a true partner for experiment, replacing dangerous and expensive lab tests in many scenarios, and permitting the creation of ultra-realistic digital twins that can be used to test the consequences of change on a manufacturing process or a new drug on a human patient. Groups who can apply such modeling capabilities have a large and growing competitive advantage over those who cannot.

While we described these three developments separately, the future lies in their ever closer integration—so that, for example, a team designing new energy-efficient materials will leverage cloud services that deliver value-added access to large datasets constructed by AI-guided simulations. What we call *discovery cloud services* underpin the whole, hosting datasets, brokering access to advanced simulation, enabling simulation, and the like. Such systems will shape essentially every aspect of the global economy over the next decades, from agriculture to environment, healthcare, manufacturing, and beyond.

**The need for a National Discovery Cloud**

Nowhere are the implications of this new computational fabric more profound than in research and education, two sectors that have long played an outsized role in sustaining US competitiveness. But to a large extent, the US research and education communities lack access to important elements of this new fabric. DOE and NSF supercomputers provide access to powerful simulation capabilities, but with access limited to small communities. With a few notable exceptions, AI-ready datasets for research use are lacking. Commercial clouds are accessible to anyone with a credit card, but there is little of the coordination needed to create nationally useful discovery cloud services. The results of these lacks and gaps are serious. Researchers who lack access to these capabilities are less competitive than their competitors, an effect that will only grow in the coming decades. Lack of access also hinders the education sector in preparing a workforce for the new jobs that will emerge in this new computationally enabled future.

---

[3] A 20-Year Community Roadmap for Artificial Intelligence Research in the US, CCC and AAAI, https://cra.org/ccc/visioning/visioning-activities/2018-activities/artificial-intelligence-roadmap/; Final Report, National Security Commission on Artificial Intelligence, https://www.nscai.gov/wp-content/uploads/2021/03/Full-Report-Digital-1.pdf

New capabilities and leadership are required if the US research and education enterprise is to effectively harness this new computational fabric for discovery, innovation, and workforce development. The challenge is to enable researchers, educators, students, and industrial collaborators to develop and use the value-added services that will underpin the society of tomorrow; aggregate the massive datasets required for AI-driven discoveries and innovation; and construct and run the simulation models used to understand future products and scenarios. Such capabilities are not provided by the ad hoc mix of campus computers, supercomputers, and commercial clouds that researchers and educators rely on today. Both new approaches and new investments are needed to provide these capabilities. Much as the establishment of the NSF supercomputer centers in the 1980s spurred US innovation in simulation science in the late 20th Century, a new **National Discovery Cloud** (NDC) is needed to sustain and accelerate innovation in digital technologies for the 21st Century.

**A National Discovery Cloud requires new capabilities**

Realizing the benefits introduced above requires the development and deployment of substantial new capabilities, such as the following.

- The definition, creation, and curation of large **reference datasets** to fuel new data-driven models of the natural world, economy, human physiology, healthcare system, manufacturing processes, etc.
- A **discovery cloud platform** to enable the collaborative development of value-added services that support NDC-powered scholarship and education
- New **educational programs and curricula** to prepare a generation for whom programming and using discovery cloud capabilities is second nature.
- Substantial **computing, storage, and network resources** to host and compute over enormous datasets and to host and operate discovery cloud services that enhance the value of datasets.
- Innovative integrations of NDC capabilities with high-performance computers, automated laboratories, and other elements of a 21st century **discovery and innovation ecosystem**.
- **Privacy and security** designed in from the beginning, rather than added post facto, and with integrated assurances and audit capabilities so that the NDC advances rather than hinders computing in the public interest.

As in prior areas in which US government investment has spurred revolutionary change, such as the Internet, digital libraries, and high-performance computing, this program will need to embed the *persistent resources* needed to support large and diverse research and education communities within an environment that encourages and supports *innovation* by those communities. Just how to strike this balance between persistence and innovation, how to enable staged design and evolution of different components, how to establish and manage the public-private partnerships that will surely be required for many of these capabilities—these are all questions that must be considered as we establish an NDC. Some of these questions were also considered in a 2018 NSF report[4].

**Discovery cloud or research cloud?**

The term "national research cloud" has been used recently by advocates for large investments in AI supercomputers to enable academic researchers to out-innovate the likes of Facebook and Google[5]. We agree that improved access to AI supercomputers is needed. However, ensuring US leadership in the 21st Century requires far more than enabling big computations by a few. AI is an important part of the 21st Century digital transformation, but only a part.

**Open Research Issues**

The vision we have described offers tremendous promise for accelerating and democratizing computing research and for advancing our economic competitiveness. At the same time, such powerful tools can prove to be double-edged swords, as we have seen with other recent developments in the field (e.g., social networking, and certain applications of AI and machine learning). Here we summarize the major issues that we believe must be addressed as steps are taken to develop and deploy a National Discovery Cloud. While some should be considered necessary conditions before certain NDC uses are enabled, in total they should be viewed as providing rich opportunities for interdisciplinary investigations (distinct positives) and not as insurmountable hurdles that render NDC infeasible.

- Substantial research takes place at the hardware and lower systems levels in the computing "stack." Many advances that we benefit from today are a direct result of academic research in these areas over the past several decades[6]. By abstracting away lower-level details of the system, cloud computing makes software developers more productive within the confines of the services supplied by a commercial cloud provider—but also cuts off the possibility of experimentation at lower levels. Indeed, commercial cloud providers may need to actively prohibit such investigations in order to ensure the security of their platforms. Accommodations must be made for computing researchers who work on hardware and lower systems levels of the stack, otherwise NDC may stunt whole lines of computing research.
- **Privacy and security**: As we have seen in recent years, the proliferation of machine learning applications combined with the data collection capabilities of social networking and smart devices has both powered remarkable advances and introduced significant new threats to our privacy. Likewise the cybersecurity of data and services remains a fundamentally unsolved problem. An NDC that makes large quantities of data available to many researchers will enable valuable scientific discoveries, but we must also be attuned to the increased risks to privacy and security. The tension between "'free and open" and "private and secure" may seem unresolvable and hence unapproachable, but it should be regarded as an extremely important area for basic research.
- **Sustainability** is an important goal for all aspects of society, and will likely be a key research area supported by NDC. Ironically, advanced AI research is nearly always characterized as the use of high powered compute farms, special purpose hardware for machine learning (GPUs), and enormous data centers. The amount of energy used to generate the experimental results needed to publish at top AI conferences has increased dramatically over the past several years. Without making other changes to the research culture, providing computing resources like this to broader segments of the research community could have a significant negative impact on the sustainability of computing. How can an NDC be organized so as to contribute to increased rather than reduced sustainability?
- More broadly, the **democratization of computing research** is a laudable goal for NDC. The unhealthy concentration of certain lines of research in large industry players in a few major metropolitan areas has also been noted elsewhere[7]. There has been a narrowing of focus across the entire field (a "chasing the leaders" effect), intense competition for certain specialities that has pulled researchers from academia to industry, and perhaps also an associated reduction in various forms of diversity. While NDC will make a substantial impact in addressing these issues, other structural changes to the research enterprise are undoubtedly necessary to overcome the full suite of challenges that are barriers toward true democratization of computing and AI research.

---

[4] J. Rexford et al. *Enabling Computer and Information Science and Engineering Research and Education in the Cloud*. National Science Foundation, 2018. https://doi.org/10.13140/RG.2.2.30487.83364
[5] https://hai.stanford.edu/national-research-cloud-joint-letter
[6] https://cccblog.org/2016/07/27/the-tire-tracks-diagram-corrected-and-humanized-by-national-academy-workshop-report/

Many open questions remain on the approaches best suited for deploying NDC. At one end of the spectrum, one or more institutions could be funded to buy, deploy, and run bespoke cloud centers, as was done for NSF supercomputer centers[8]. At the other end of the spectrum, contracts could be established with private-sector cloud providers to give researchers access to computation and data storage at little or no charge to the researchers. Intermediate solutions include having a private-sector cloud provider manage cloud instances at a funded center. Tradeoffs abound. Bespoke centers can enable access for research into providing cloud infrastructure—low-level software, firmware, and even hardware—while the private-sector cloud alternatives offer great efficiencies by leveraging the billion-dollar investments that major cloud providers make. Scientists may require specialized capabilities and support for their unusual applications and communities, both for software development and for purposes such as problem detection and optimization. Careful thought is needed for managing appropriate use of valuable data in centers used by many researchers, including for low-level research. The tension of these and other requirements may preclude a good "one size fits all" solution and necessitate two or more varied deployments.

*The views expressed in this whitepaper are those of the authors and do not necessarily reflect the views of their employers.*

---

[7] https://www.nscai.gov/2021-final-report/
[8] https://journals.sagepub.com/doi/pdf/10.1177/109434209100500401

*This material is based upon work supported by the National Science Foundation under Grant No. 1734706. Any opinions, findings, and conclusions or recommendations expressed in this material are those of the authors and do not necessarily reflect the views of the National Science Foundation.*